\documentclass[journal,transmag]{IEEEtran}

\usepackage{amssymb}
\usepackage{graphicx}
\usepackage{caption}
\usepackage{subcaption}
\usepackage{algorithm}
\usepackage{algorithmic}
\usepackage{epstopdf}
\usepackage{color}
\usepackage{cite}

\begin{document}

\bibliographystyle{unsrt}

\title{Evaluating Reputation Management Schemes of Internet of Vehicles based on Evolutionary Game Theory}

\author{\IEEEauthorblockN{Zhihong Tian\IEEEauthorrefmark{1},
Xiangsong Gao\IEEEauthorrefmark{2},
Shen Su\IEEEauthorrefmark{1},
Jing Qiu\IEEEauthorrefmark{1},
Xiaojiang Du\IEEEauthorrefmark{3},
and Mohsen Guizani\IEEEauthorrefmark{4}}

\IEEEauthorblockA{\IEEEauthorrefmark{1}The Cyberspace Institute of Advanced Technology, Guangzhou University, Guangzhou, China\\\IEEEauthorrefmark{2}Institute of Computer Application, China Academy of Engineer Physics, Mianyang, Sichuan, China\\\IEEEauthorrefmark{3}Dept. of Computer and Information Sciences, Temple University, Philadelphia PA 19122, USA\\\IEEEauthorrefmark{4}Dept. of Electrical and Computer Engineering, University of Idaho, Moscow, Idaho, USA}
\thanks{Corresponding author: Shen Su and Jing Qiu}}

\maketitle

\begin{abstract}
Conducting reputation management is very important for Internet of vehicles. However, most of the existing researches evaluate the effectiveness of their schemes with settled attacking behaviors in their simulation which cannot represent the scenarios in reality. In this paper, we propose to consider dynamical and diversity attacking strategies in the simulation of reputation management scheme evaluation. To that end, we apply evolutionary game theory to model the evolution process of malicious users' attacking strategies, and discuss the methodology of the evaluation simulations. We further apply our evaluation method to a reputation management scheme with multiple utility functions, and discuss the evaluation results. The results indicate that our evaluation method is able to depict the evolving process of the dynamic attacking strategies in a vehicular network, and the final state of the simulation could be used to quantify the protection effectiveness of the reputation management scheme.
\end{abstract}

\begin{IEEEkeywords}
reputation management scheme, Internet of vehicles, evolutionary game theory, malicious users, utility function
\end{IEEEkeywords}

\section{Introduction}

The rapid development of mobile Internet gives us an opportunity to enjoy a better life when traveling with vehicles. With on board units (e.g. smart phones), vehicles are able to share information over the Internet at real time, and learn bigger visions all over the road map to make better travelling decisions. Within the last few years, a bunch of business has arisen for intelligent traveling, e.g. intelligent traffic information service \cite{al2015intelligent}\cite{vorona2016system}, taxi booking apps \cite{qiao2015efficient}, and unmanned vehicles \cite{pestana2014computer}\cite{tokekar2016sensor}. However, the growing business of intelligent traveling also brings growing security problems since malicious behaviors are turning to be more threatening. In many fields, different problems and related research results are proposed. Such as the disclosure of privacy data \cite{yu2018data}\cite{wang2018privacy}, the real-time nature of event data \cite{tian2018real}, the attack against peer-to-peer networks \cite{tan2018towards} and detection or forensics of these attacks\cite{jiang2018deep}\cite{zhihong2015transductive}\cite{zhihong2014digital}, etc. One of the biggest problems for the business runner is to identify the malicious (or bad) users of the intelligent service users. For example, in a traffic information service, when the malicious vehicles broadcast fraud messages, the other vehicle would be misled by the fraud message. In a taxi booking application, if an application driver always refuses to serve for assigned orders, the passengers of such orders would definitely have bad experience on such taxi booking applications. Herein, a reputation management scheme of such business over the Internet of vehicles is very important for improving its service quality and service experience.

The reputation management scheme (also noted as trust model) of vehicular networks has been extensively studied within the last decade \cite{chen2018trust}. Generally, a reputation management method sets up a global or private credit for each entity, data, or both. It also adjusts the credit according to the dynamic evaluation of the entity or data by punishing the misbehaviors \cite{ts1}\cite{ts2}. When it comes to the evaluation of the effectiveness of a trust model, the behavior of the attackers are always predefined, and all attackers act in the same way, e.g. ``In our experiments, the sender weight factor is set to 2 to double the weight of a sender in message evaluation'' \cite{chen2010trust}. As an experiment simulation for qualitative analysis, we think this kind of assumption is okay. However, we doubt the rationality of such assumptions when it comes to the practical use.

First, due to the different backgrounds and habits, the attackers are more likely to take different attacking strategies which leads to the attacking behaviors varying in a wide range, and happening simultaneously. Therein, the attackers in the same simulation should take various actions to be closer to the de facto practice.

Second, we should not assume all attackers are wise enough to attack. Since a malicious user may be blind to the overall reputation management scheme and the behaviors of the other users in the same intelligent travelling service, it is really hard to make a smart choice among all the possible attacking strategies. In a real case, we believe the story is like this: the original attacking plan is not profitable enough to the attacker and the attacker learns from the previous experience according to the punishment of the vehicular network and improves their plan to attack wisely.

According to the above discussion, the simulation of a reputation management scheme evaluation should involve malicious vehicles with a set of dynamic and diverse strategies, and the current common practice of reputation management scheme evaluation still has a long way to the end of practical use. To fill the gap between the industrial requirements and research output, we propose to apply evolutionary game theory in the evaluation simulation of a reputation management scheme of vehicular.

Our idea is to simulate the strategy selections of the malicious vehicles in the evaluation process following the evolutionary game mode. We initialize the population of malicious vehicles with randomly selected strategies in a wide range, and give the malicious vehicles opportunities to evolve by making a heuristic change according to the feedback of the reputation system. In general, the malicious vehicles would converge to a final state which is an optimal choice for all.

Our contribution in this paper include 3 folds:

1) We take a first step for the simulation that attackers in reputation management system of a vehicular network would take diverse and dynamic strategies, which provides an evaluation scenario much closer to the real world, and enables the research community to conduct systematically and evaluate a reputation management scheme.

2) We propose to apply evolutionary game theory in the evaluation process of reputation management schemes, and discuss the detailed simulation scenarios of game modeling.

3) We apply our evolutionary game method to an example reputation management scheme with various attacking targets. The evaluation results suggests that our method is able to depict a detailed evolution process. Since the evolution converges to the state that all malicious entities get to its optimal strategies, our evolutionary game method is also able to provide a quantitative evaluation about the effectiveness of a reputation management scheme.

The rest of the paper is organized as following. First, we introduce the related works in Section 2. Then we formally define our problem in Section 3 and describe our evolutionary game model and evolution progress together with a running example in Section 4. In Section 5, we apply our evaluation model on our example reputation method to describe its effectiveness in reality. Finally, we conclude our paper with feature works in Section 6.

\section{Related Works}

\subsection{Problems in VANET and Trust Models}

VANET can be built on V2V and V2I networks, most of which use wireless communication technology such as WiMAX or IEEE 802.11p \cite{v2inetworkscompare}. The research community has long noticed possible security issues in VANET, and \cite{vs1}\cite{vs2} has investigated and summarized these security issues. The trust problem, how to minimize the damage caused by false information transmitted by dishonest communications entities, has become the focus of research. The work related to solving the trust problem has been fully surveyed in \cite{ts1}\cite{ts2}. Moreover, \cite{vts} has researched and summarized the trust models in VANET. The trust models in VANET can be divided into three categories: entity-oriented and data-oriented, in addition to some work that combines the characteristics of both types of models.

In the entity-oriented models, \cite{eom1}\cite{eom2} evaluates the trust of the entity. The environmental factors, other entities that have communicated with the entity, and the historical behavior of the entity all have an impact on its trust evaluation. As for the data-oriented model \cite{dom1}, instead of the entity itself, the data sent by the entity is given credibility. By defining an appropriate trust metric, an entity does not focus on other entities, but rather measures the level of trust in the data it receives.

There are still some problems with these two models. Entity-oriented models are sensitive to the number of sources of information. The trust relationship between entities under the data-oriented model will never change - this is the inefficient \cite{vts}. \cite{com} was proposed to overcome these problems. Under the model, the entity simultaneously measures the credibility of the received data and the trust evaluation of the communication opponents from other entities. Several papers \cite{du15}\cite{du16}\cite{du17}\cite{du18}\cite{du19} have studied related IoT security issues.

\subsection{Evolutionary Game Theory}

At the beginning of game theory, it was a branch of mathematics in service economics \cite{laozuzong}. However, with the development of game theory, computer science and engineering have gradually begun to apply game theory to solve problems. \cite{gs1} summarizes the existing game theory application results in the field of network security and proposes a classification method. \cite{gs2} investigates the existing game theory application results in the field of wireless network engineering. Especially on the issue of distributed denial of service attacks, game theory has been widely used \cite{gapp1}\cite{gapp2}\cite{gapp3}.

The solution under the traditional game theory framework, Nash equilibrium \cite{zuzong}, has a strong assumption: all game players are absolutely rational. Researchers have recognized this problem for years and tried to overcome \cite{br1}\cite{br2}. In the field of information security, \cite{evoapp1} proposes the two characteristics of bounded rationality and repeated game in network confrontation game, which is the premise of evolutionary game theory. On this basis, the application of evolutionary game theory in economics and networks is comparatively studied. In the field of wireless network engineering, \cite{evoapp2} proposes a dynamic network selection scheme based on evolutionary game theory under heterogeneous networks.

Due to the convergence of evolutionary game theory and VANET, there have been some research results in this field. \cite{evoapp3} establishes an evolutionary game model for VANET and studies the issue of incentive cooperation in VANET. They believe that the cooperative behavior in VANET will not spread unconditionally. And \cite{evoapp5} also studied the cooperative behavior in VANET. They found through simulation of the evolutionary game model that the cooperation rate is related to the energy consumption factor. And compared to all vehicles choose cooperation, a certain proportion of partners can achieve the same information dissemination effect, while saving some bandwidth. \cite{evoapp6} proposes a routing algorithm based on evolutionary game theory to solve the selfish problem in VANET. \cite{evoapp4} studied the security protection scheme in VANET under the framework of evolutionary game theory, and established a generalized evolutionary game model with versatility. They use threats posed by malicious attackers as input to optimize the deployment of the security protection scheme. \cite{evoapp7} proposes an evolutionary game model for the RSU bandwidth resource competition in V2I networks, and its evolutionarily stable strategy is considered to be the best solution for this competition. They also simulated evolutionary games, showing the evolutionary process and the impact of evolutionarily stable strategy on network utilization efficiency.

Our work is not to propose a new trust management model or security protection scheme, but to find a solution that can effectively and rationally evaluate a trust management model under the framework of evolutionary game theory.

\section{Problem Statement}

In general, our goal is to find the most effective protection solution deployment decision for fraud in the connected vehicles’ network. In this part of the following, we will use the ordinary V2I network and a simple reputation calculation model as the research object to illustrate the basic assumptions and problems.

\subsection{V2I Network}

In our V2I network, vehicles communicate with the central server to exchange traffic event messages. These traffic events are events that occur randomly on the road network that may affect the vehicle's travel (such as traffic jams, road construction, traffic accidents, and road conditions, etc.). Vehicles traveling on the road have the ability to sense these traffic events. Once the vehicle senses the presence of a traffic event, it can form an event message and submit it to the central server. The central server is responsible for managing the vehicle information and the messages they submit. In addition, the central server also broadcasts a list of traffic events on the road network to the vehicle to assist in driving. Finally, the number of vehicles traveling on this road network can vary from moment to moment. In other words, new vehicles can be added to the network at any time, and vehicles can leave the network at any time. However, we assume that the total number of vehicles traveling on the road network is relatively stable over a certain period of time. Similarly, the proportion of dishonest vehicles in the overall vehicle is relatively stable.

\subsection{Deceptive Behavior}

As we mentioned in the first (second) section, deceptive behavior can have a negative impact on this connected vehicles’ network system. Specifically, deception refers to a dishonest vehicle submitting its counterfeit false (non-existent) traffic event message during the driving process. The strength of submitting false messages can be called deception intensity. We define this as the average rate at which dishonest vehicles submit false traffic event messages to the central server. That is, the number of false traffic events messages submitted to the central server per unit time. For example, in hours, a dishonest vehicle submits ten false traffic events messages to the central service within one hour, and the deception intensity is ten. In this paper, we take 5000 seconds as a unit of time. The deception intensity of dishonest vehicles is:
\[fr \in \{ x \mid 1 \leq x \leq 100, x \in Z \}\]

Obviously, for a single dishonest vehicle, choosing a different level of deception intensity can have varying degrees of negative impact on the system. For the entire dishonest vehicles with uncertain totals in the network, regardless of whether they are connected, they can choose different deception intensity at the same time. Different distribution of population with different deception intensity make different overall negative impact on the system. In addition, the distribution of deception intensity can even change at any time. We cannot simply assume that dishonest vehicles are unanalyzable. As time progresses, dishonest vehicles should generally be able to change the distribution of deception intensity in order to achieve greater negative impact.

\subsection{Trust Management Model}

This paper takes a simple trust management calculation model we designed as a research object, which combines the characteristics of both entity-oriented and data-oriented models. Under this model, vehicles and traffic event messages have their own reputation values.

For the reputation value of the vehicle, the system gives each vehicle in the network a reputation value. If the vehicle's reputation value drops to zero, then the vehicle will be judged to be dishonest by the system. Dishonest vehicles will be removed from the network. As for the reputation value of traffic event messages. When a traffic event message is submitted, the system assigns the reputation value of its submitted vehicle to the traffic event message as an initial value. If the reputation value of the traffic event message becomes zero, the message is determined to be a false message. False messages will be removed from the broadcast list.

In terms of reputation calculation for a traffic event message, whenever the system receives a report that does not exist, the reputation value of the message is reduced by one unit. And for a vehicle, whenever a traffic event message submitted by it is removed, the system will perform a punitive reduction on the reputation value of the vehicle.

Obviously, there are some parameters that need to be preset when deploying such a protection scheme. The method of penalizing vehicle reputation values (such as linear decline, exponential decline, and logarithmic decline, etc.) is a typical option. The protection effect produced by choosing different punishment methods is naturally different.

\subsection{Summary}
In summary, we found that there are some options in the deception of this connected vehicles’ network that can affect the outcome where these choices can all be dynamic. For the distribution of deception intensity in the population, we aim to analyze the law of change and find the optimal distribution and its maximum benefit (for dishonest groups). On this basis, we will further analyze the impact of different trust management model parameters on these optimal distributions and maximum benefits. This indicates the optimal deployment decision for the protection scheme.

\section{Evolutionary Game}

Evolutionary games occur between individuals in the population. At a point in time, the distribution of decisions on the population is called a state. The evolutionary game is a dynamic system in the mathematical sense. The object of its study is the development of the trajectory of this state in the state space, which is characterized by a differential equation. This differential equation is called the replicator equation in the evolutionary game. According to evolutionary game theory, there are steady states in the dynamic system \cite{thomas1984com}. These steady states are called evolutionary stability strategies or evolutionarily stable sets as solutions to evolutionary games. Often these evolutionary steady states will bring higher utility to the overall population. In this section, we will first introduce our evolutionary game model through the four elements of evolutionary games. We will then explain the details of the evolution process.

\subsection{Game Definition}

This evolutionary game can be formalized as a four-tuple: $(P,S,A_{t},U)$.

\begin{enumerate}
\item
$P$ represents the player in the game. For the problem of deceiving in the connected vehicles’ network, we define the player of the evolutionary game as a whole population of dishonest vehicles. In all dishonest vehicles, the decision of every dishonest vehicle is influenced by other dishonest vehicles. As we mentioned in the third section, the total number of players is macroscopically stable.

\item
$S$ represents the player's strategic space. The strategy for dishonest vehicles is the deception intensity $fr$ we mentioned in the third part. That means the strategic space S is a discrete collection of 100 elements.
\[S = \{ x \mid1 \leq x \leq 100, x \in Z \}\]

\item
$A_{t}$ represents the distribution of decisions in the population at time t. Different individuals in the population can choose different strategies, and all individuals who choose the same strategy constitute the group that chooses the strategy. The proportion distribution of all groups in the population as an overall decision. This overall strategic space is the state space of the evolutionary game dynamic system. The individual's strategic space composition determines that the state space of the evolutionary game dynamic system is a regular simplex in 100-dimensional Euclidean space.
\[A_{t} \in \{ (a_{1}, a_{2}, \cdots , a_{i}, \cdots , a_{100}) \mid a_{i} \in [0, 1], \Sigma a_{i} = 1 \}\]

\item
$U$ represents the individual's utility function. We define it as the sum of the durations of all false event messages submitted by the vehicle. The duration of the message represents the time elapsed since the message was submitted to be removed. On the definition of individual utility, there is further utility of the group $U_{i}$. It can be simply defined as the sum of individual utility in the group whose decision is $i$.
\end{enumerate}

\subsection{Evolution Process}

In nature, the evolution of a population is reflected in the process of reproduction and natural selection. In our evolutionary game model, natural selection is reflected in the fact that dishonest vehicles are removed from the network by the system as dishonest. And reproduction is reflected in the fact that when the vehicle is removed from the network in the population, we call it elimination, the new vehicle will join the population and enter the network. The practical significance of this is that the owner of a dishonest vehicle may re-enter the network and continue to attack by replacing his account on the network.

Evolution takes place on the strategic choice of newly joined vehicles. According to our replicator equation \cite{thomas1985sim}:
\[\dot a = a_{i} [U_{i} - \sum_{j=1}^{100} a_{j}U_{j}]\]

The algorithm for the newly joined vehicle selection strategy is as follows:

\begin{algorithm}
\caption{Strategy selection algorithm}
\label{alg:ssa}
\begin{algorithmic}
\STATE observe utility $U_{list}$  for all strategy
\STATE observe $U_{overall} = \sum_{j=1}^{100} a_{j}U_{j}$
\FOR{all $U_{i} \in U_{list}$}
\STATE $list_{predominant} \leftarrow decison i$
\STATE $decison \leftarrow RandomSelect(list_{predominant})$
\ENDFOR
\STATE Join the group and execute the attack strategy
\end{algorithmic}
\end{algorithm}

The above algorithm assumes that dishonest vehicles can always choose a better strategy. It may be an overestimation of the attacker in the actual scene. In this regard, we can define the evolution ability factor as an accuracy that affects the attacker's choice of a new strategy. This factor can be affected by many factors, such as the attacker's analytical ability, the attacker's intelligence gathering ability, the degree of information exchange between a particular attacker and other attackers and whether there is a cooperative relationship between the attackers. When the evolution ability factor is 1, the above algorithm is established. This represents one of the worst situations in both theoretical and practical scenarios.

\subsection{Evolution Example}

Next, we use an evolutionary game on a simple 3D strategy space as an example to illustrate the evolution of the state of evolution. Let us assume a scenario like this:

\begin{itemize}
\item The total number of dishonest vehicles is 100
\item The initial state of dishonest vehicles, ie the population strategy distribution in the population, as shown in figure 1, the three strategies are 10, 20, 30 in order.
\end{itemize}
 
\begin{figure}[h!]
\includegraphics[width=\linewidth]{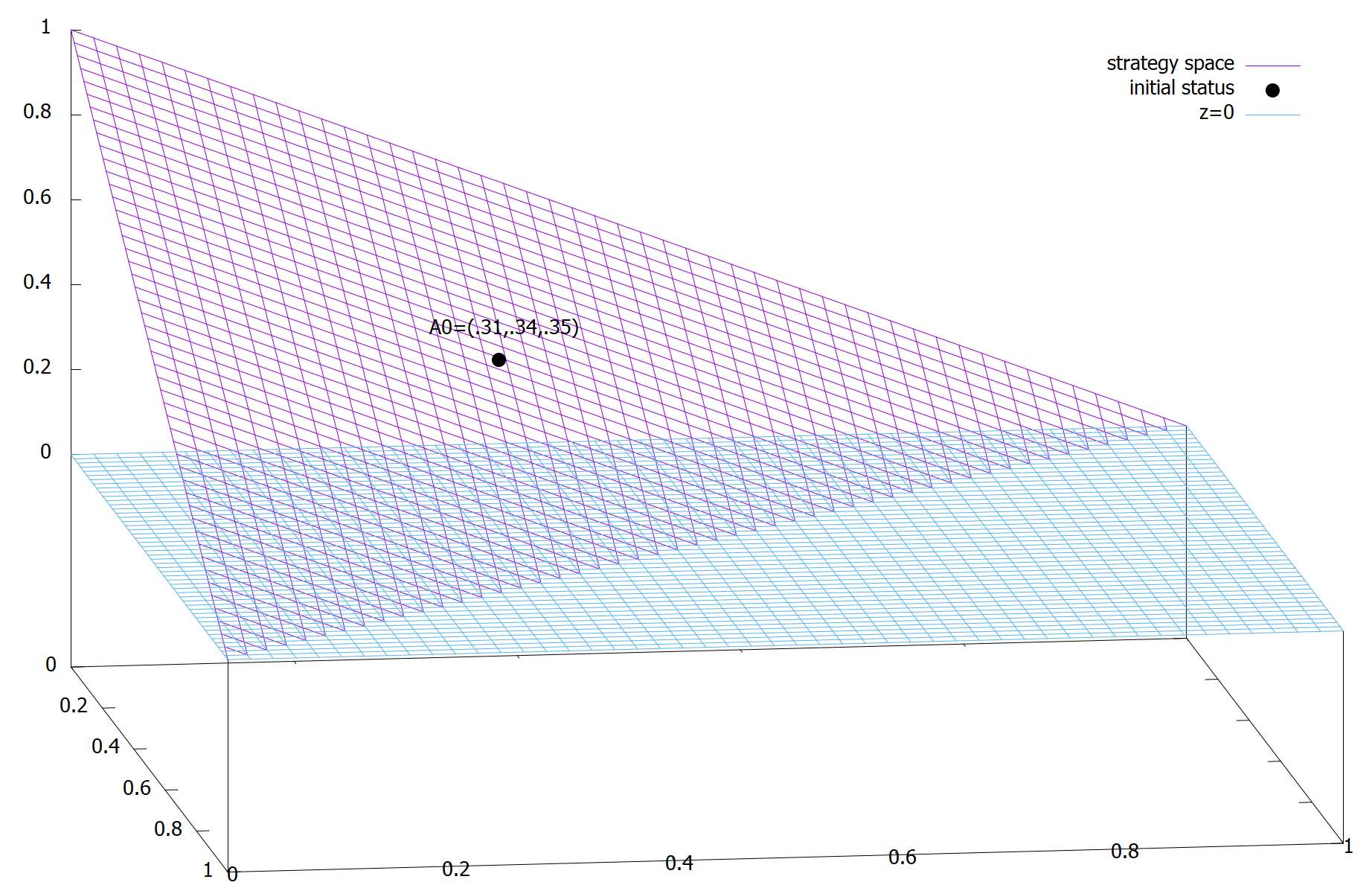}
\caption{The status space and initial status}
\label{fig:example-init}
\end{figure}

As time goes by, vehicles will gradually be removed. At this point, the newly joined vehicles will need to re-select their strategy. As shown in the table 1, some vehicles in the Strategy 10 group were eliminated between time 1004 and 1005. New vehicles want to join the network to continue the attack. These new vehicles need to choose their own attack strategy according to the strategy selection algorithm. At this time, only the utility of Strategy 20 and Strategy 30 (16171 and 15326) is higher than the overall utility (13992.247). As a result, these new vehicles have chosen from these two strategies. Note that the new vehicle does not directly select the strategy of the most effective group. This is because in evolution, the utility of the group is related to the population. This means even if the group's utility is the highest, it does not mean that the corresponding strategy is the best. It is extremely difficult for individuals to choose the best strategy without collusion.

\begin{table}[h!]
\label{tab:example-tab}
\begin{tabular}{|c|c|c|c|}
\hline
Time & Distribution in population & Utility of groups & Overall utility \\ \hline
1004 & 0.19, 0.47, 0.24 & 5356, 16142, 15319 & 13645.489 \\ \hline
1005 & 0.16, 0.48, 0.25 & 5372, 16171, 15326 & 13992.247 \\ \hline
1006 & 0.11, 0.49, 0.29 & 5383, 16200, 15333 & 14580.562 \\ \hline
\end{tabular}
\caption{Group distribution status and utility between time 1004 and 1006}
\end{table}

In the long run, the overall strategic distribution of the group will start from the initial state, draw an evolutionary trajectory in the state space, and finally stabilize in the evolutionary stable state, as shown in the figure 2.

\begin{figure}[h!]
\includegraphics[width=\linewidth]{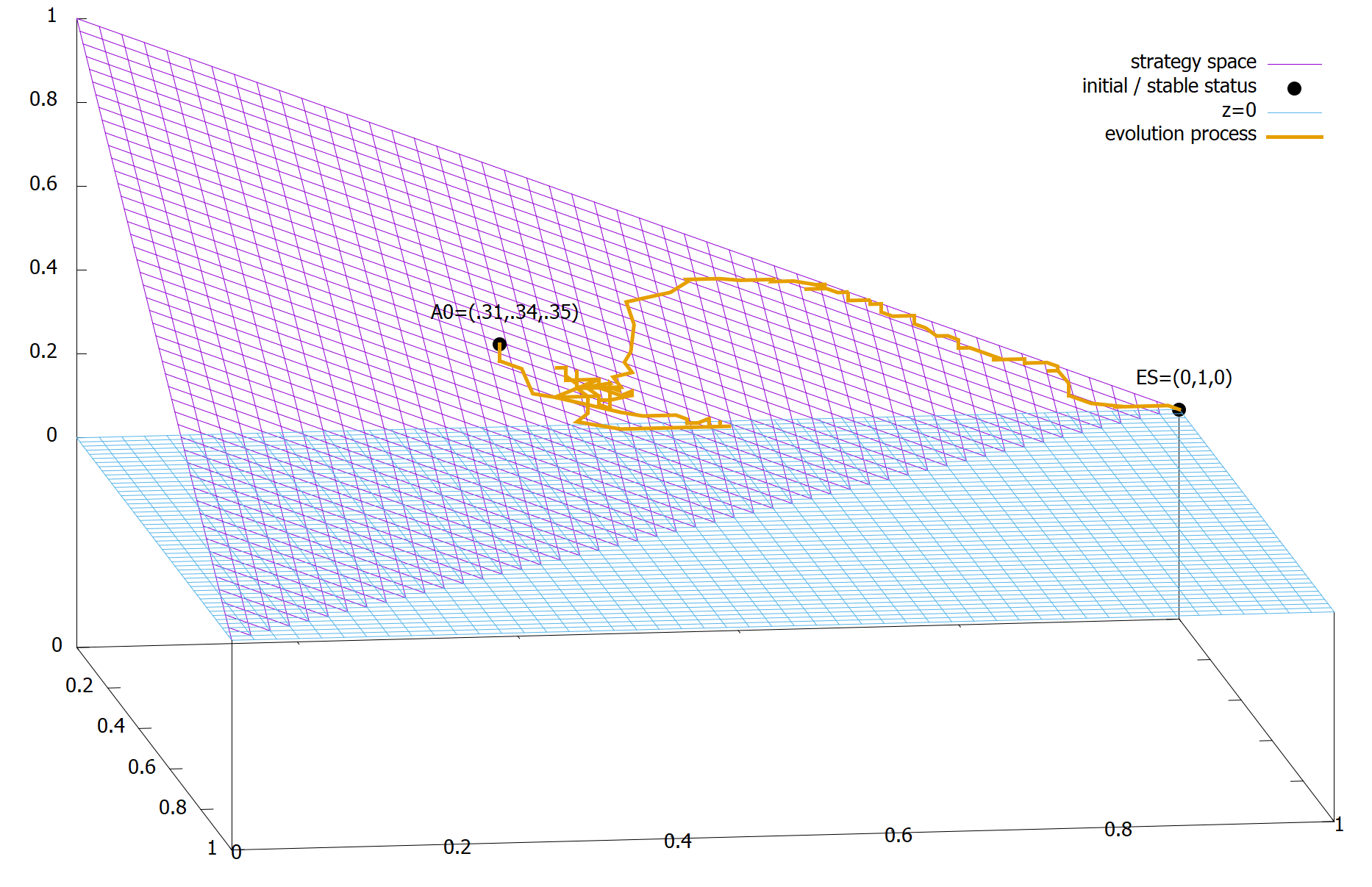}
\caption{Evolutionary trajectory}
\label{fig:example-pro}
\end{figure}

\section{Evaluation}

In this section we will show some simulations of the aforementioned evolutionary game. The simulation will mainly show the evolutionary process and steady state under different protection scheme parameter settings. In addition, we will use the average growth rate of overall utility as a measure to show the level of damage that dishonest vehicles can cause to the network in a steady state of different scenarios. The higher the average growth rate, the more dishonest vehicles can have a greater negative impact on the network at this moment.

\subsection{Experiment Setup}

Our road network, including vehicles and events, is illustrated in Figure 3. We simulate the driving process of the vehicle in the form of time slicing. The parameters in the entire road network of the road are as follows:

\begin{itemize}
\item
The size of the road network is $87 \times 87$ blocks. The length of the block is $1 km$, and the road network size without road width is $87 \times 87 = 7569 km^{2}$
\item
The vehicle travels at a constant speed of $36 km/h$. Driving mode is random walk, but will not turn around
\item
The total number of vehicles is $100,000$
\item
Among them, dishonest vehicles account for one thousandth, that is, 100 vehicles
\end{itemize}

\begin{figure}[h!]
\includegraphics[width=\linewidth]{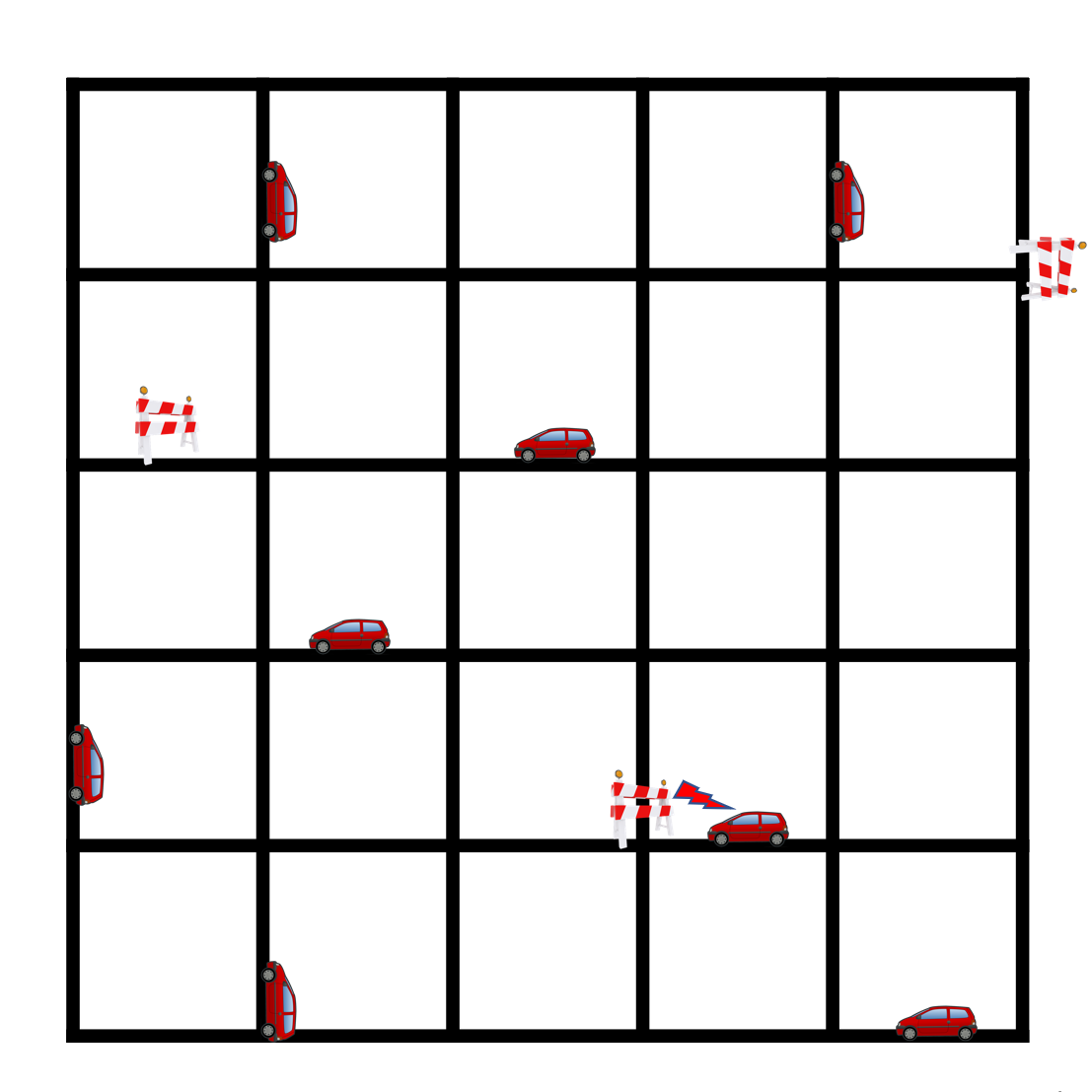}
\caption{Road network diagram}
\label{fig:road-net}
\end{figure}

In addition to all of the following simulations, the initial strategic choices for the 100 dishonest vehicles were evenly distributed. That is to say, 100 vehicles have selected the deception intensity of 1 unit to 100 units as their initial decision.

Finally, we use one of the parameters in the protection scheme, the initial reputation value assigned to the vehicle, as a strategy for the trust management model. We will also observe the effects of different parameters on the evolutionary process.

\subsection{Simulation}

\begin{figure*}
\begin{subfigure}[b]{0.333\linewidth}
\includegraphics[width=\linewidth]{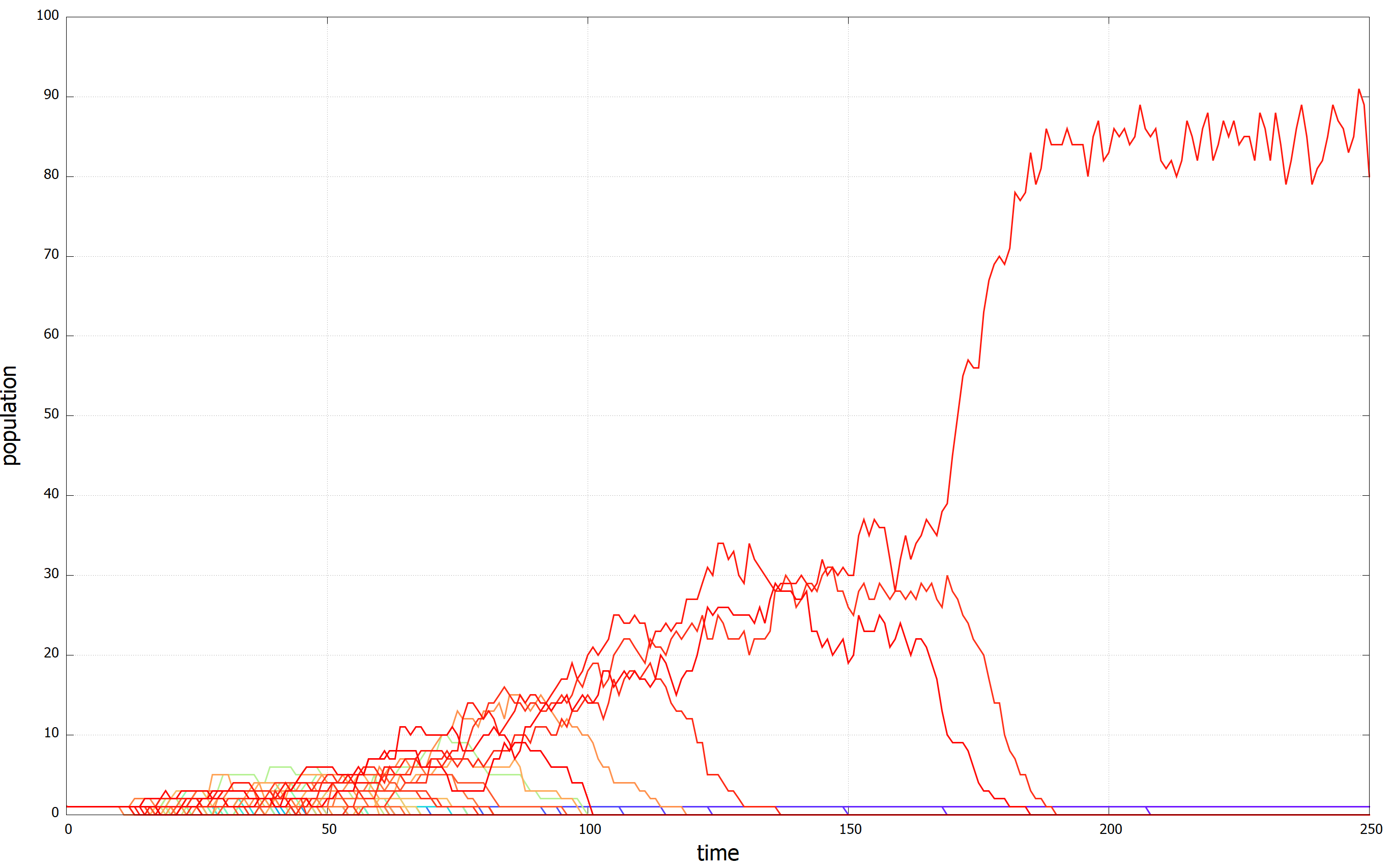}
\caption{Initial reputation value is 1. The pure strategy in steady state is strategy 97}
\end{subfigure}
\begin{subfigure}[b]{0.333\linewidth}
\includegraphics[width=\linewidth]{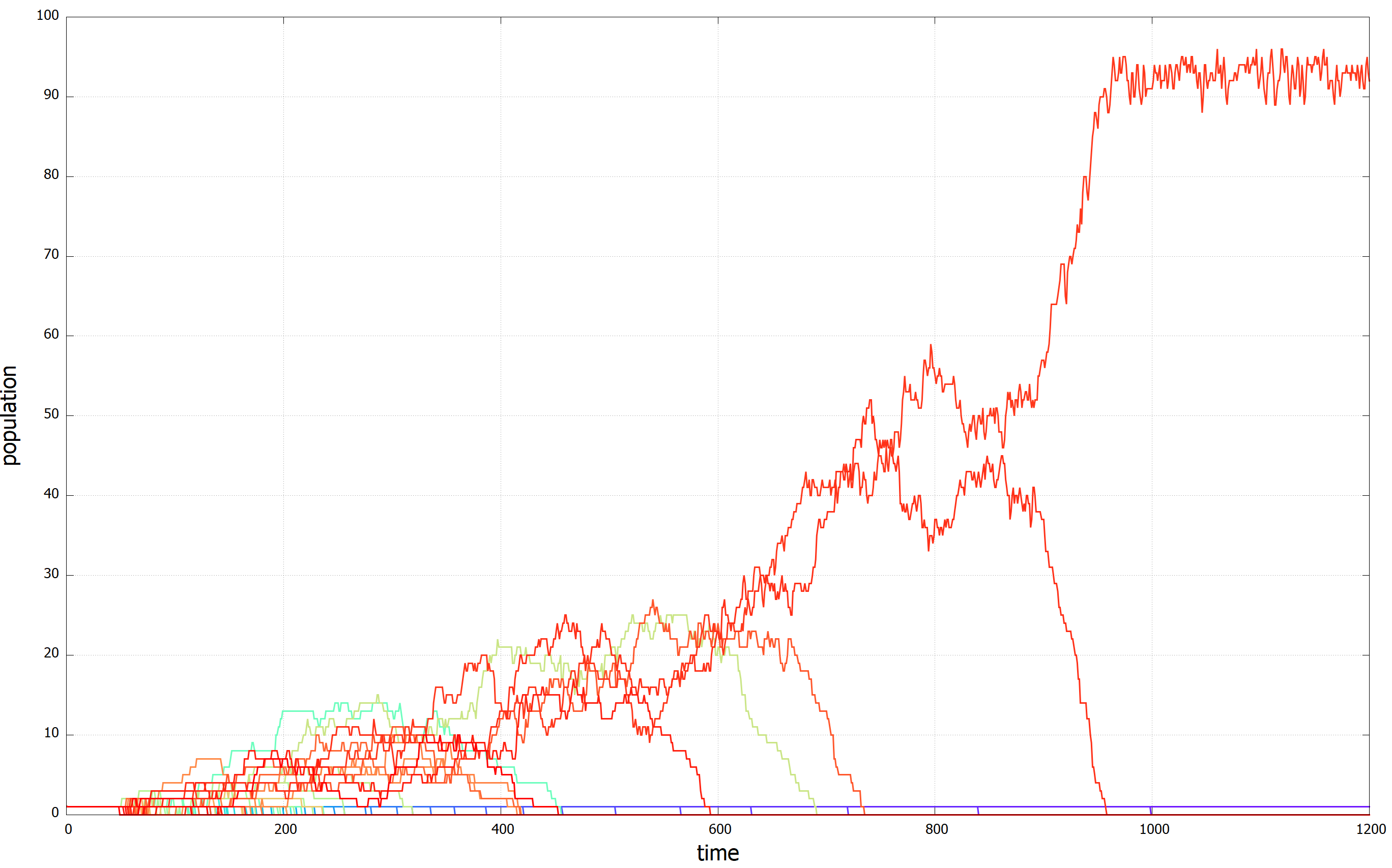}
\caption{Initial reputation value is 5. The pure strategy in steady state is strategy 93}
\end{subfigure}
\begin{subfigure}[b]{0.333\linewidth}
\includegraphics[width=\linewidth]{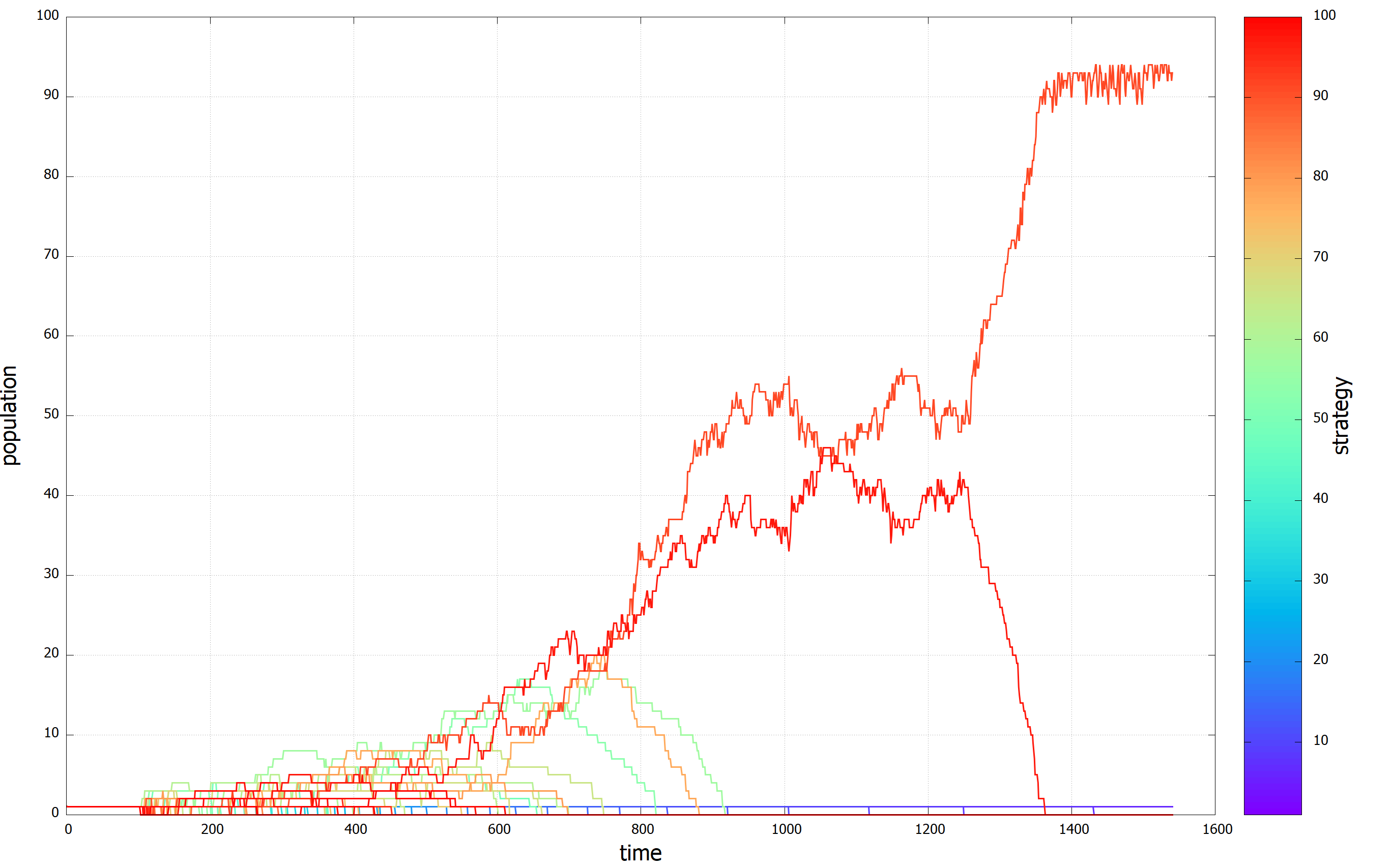}
\caption{Initial reputation value is 10. The pure strategy in steady state is strategy 91}
\end{subfigure}
\caption{Changes in population strategy distribution over time in different environments}
\label{fig:stat-com}
\end{figure*}

\begin{figure*}
\includegraphics[width=\linewidth]{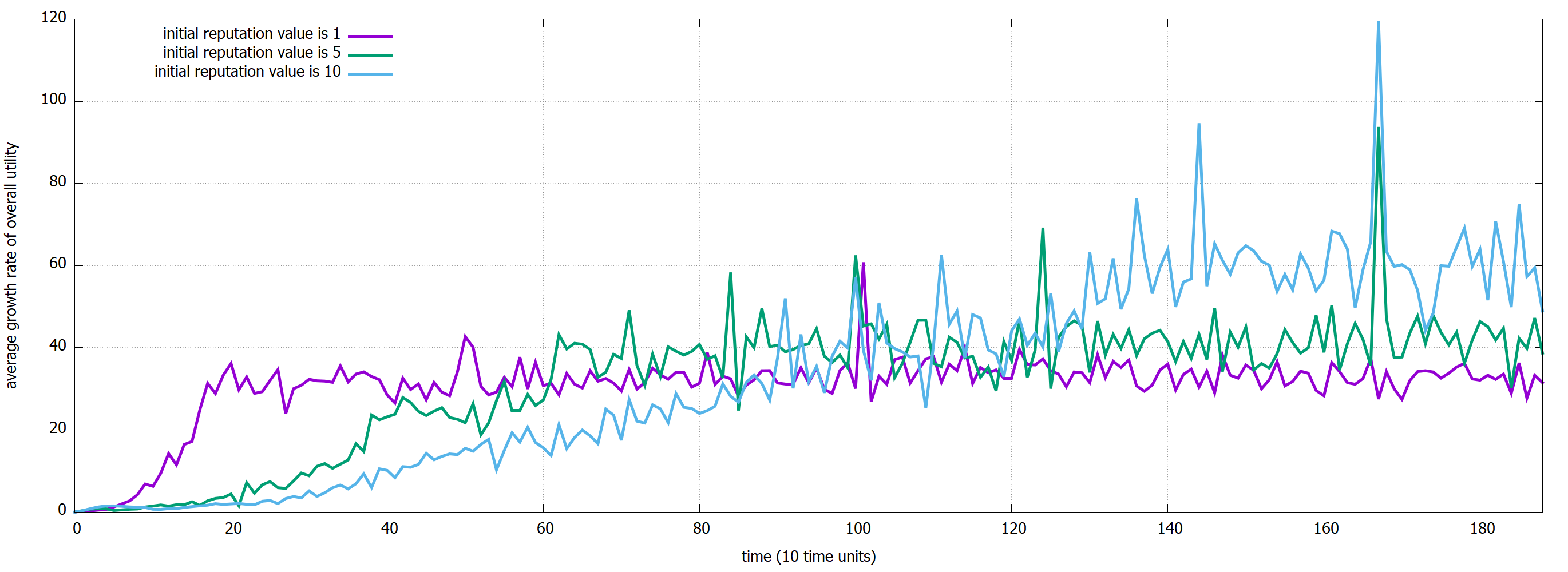}
\caption{The average growth rate of overall utility over time}
\label{fig:metric}
\end{figure*}

\begin{figure*}
\includegraphics[width=\linewidth]{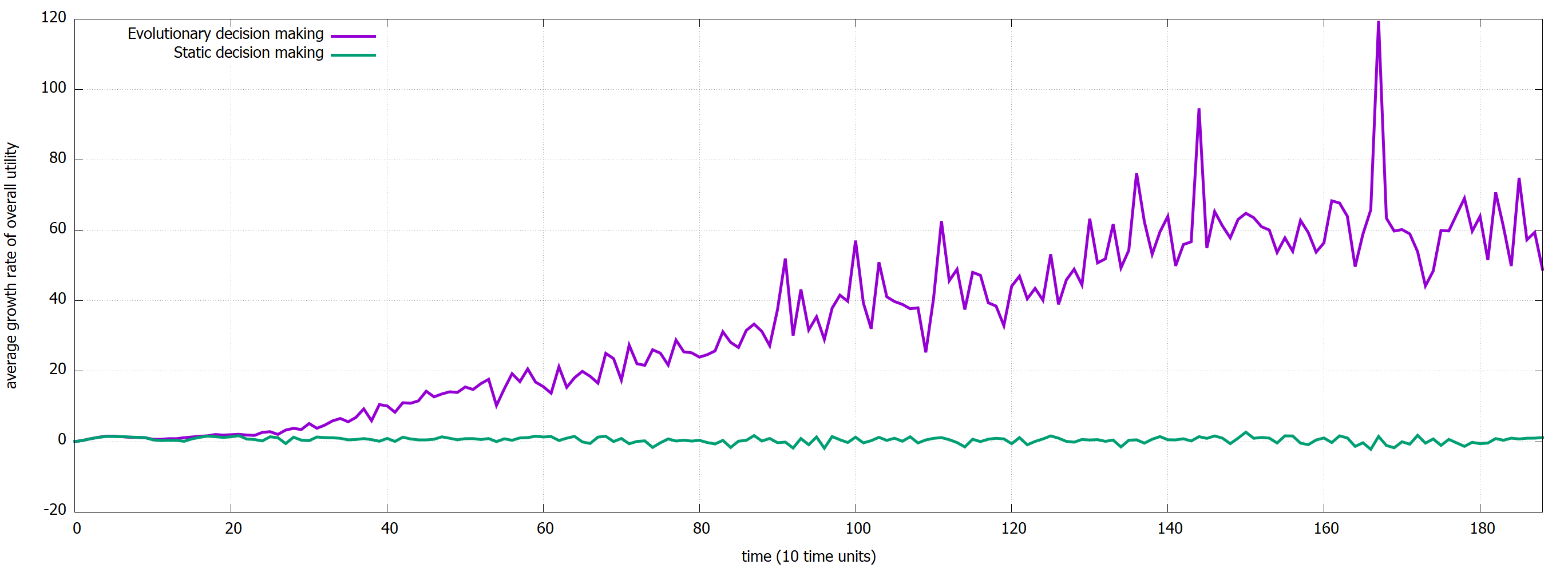}
\caption{The average growth rate of overall utility over time, with initial reputation value is 10.}
\label{fig:vs}
\end{figure*}

Figure 4 shows the population strategy distribution over time under three different trust management model settings. Lines drawn by different colors represent different strategic groups. As we mentioned earlier, 100 dishonest vehicles have chosen a deception intensity of 1 to 100 units as an attack strategy. A total of 100 strategies formed 100 groups. We can observe that during the evolution process, some groups disappeared because their utility could not keep up with the overall average. For example, in the vicinity of the 120th time unit in Figure 4.a, the group that selected Strategy 63 began to die out. Until the 150th time unit, this group eventually became extinct. On the contrary, other groups have grown stronger due to however more efficient their output may be. For example, in Figure 4.a, the size of the group that selected Strategy 97 is constantly expanding. Eventually it began to replace the group that selected Strategy 86 in the vicinity of the 175th time unit and thus rose sharply. In the end, the population strategy distribution tends to be a stable state of pure strategy. In Figure 4.a, the pure strategy in this steady state is Strategy 97. We observed that the population of this group was oscillating and less than 100 in steady state. This is because this stability is dynamic. Dishonest vehicles in the network are submitting false event messages all the time, and the system is also removing the discredited vehicles identified from the network all the time. In this way, the dishonest vehicles in the group will die at any time and will be added at any time. But supplements always lag behind death. This causes the group's population to always oscillate below 100 under steady state conditions.

Compared with different environments, that is, different protection scheme deployment strategies, the final pure strategy stable state is generally within a more aggressive range. But obviously, their convergence time is not the same. Simply put, the higher the system's tolerance for errors, the longer it takes to converge. In our opinion, this is because the high tolerance of the system leads to a decline in the speed of group replacement. In particular, high tolerance results in a single dishonest vehicle being able to survive longer in the network. Groups are more likely to maintain their decisions. Conversely, low tolerance makes dishonest vehicles removed faster. Groups are replaced more frequently. Naturally, the time it takes for evolution to reach stability is shorter. This also explains the reason why the extremely conservative strategy groups in Figure 4 can survive for a long time, and their extremely conservative strategies make them difficult to replace.

As we mentioned before, The average growth rate of overall utility is used to assess the ability of dishonest vehicles to cause damage to the network as a whole. Figure 5 shows the changes in network damage levels over time by dishonest vehicles. As the evolution progressed, the overall level of damage done to the network gradually increased. This value also tends to stabilize when the population strategy distribution is finally stable. Compare the level of damage that can be caused by the final stable in different environments. Obviously, for this simple trust management model, higher the system's tolerance for errors, the higher the level of damage that dishonest vehicles cause to the network.

Figure 6 shows the gap in the overall utility of the group decision-making method under the guidance of evolutionary games and the completely static decision-making method. The gap between the two is very obvious. Compared to the evolutionary game method, the static attack strategy seems to be unsuccessful. From the beginning to the end, the overall utility has barely increased.

\section{Conclusion}

In this paper, we discuss the feasibility of an applied scenario with dynamic and diverse attacking strategies in the evaluation of reputation management schemes of Internet of Vehicles. To that end, we propose an evolutionary game theory based solution which initializes the attacking plans with random choice, and conducts attacking plans evolution with detailed simulation consideration. With application on our example reputation management system, we manage to depict the evolution process of the attacking scenarios which converges to a situation that most of the malicious vehicles finally get to its optimal choice, and simulation result could also be used to quantify the effectiveness of a reputation management scheme of a vehicular network.

\section*{Acknowledgement}

This work is supported by the National Natural Science Foundation of China under Grant No.61871140 and No.61572153. and the National Key research and Development Plan (Grant No. 2018YFB0803504).

\bibliography{paper}

\end{document}